\documentclass[3p,onecolumn,10pt]{elsarticle}

\usepackage{placeins}
\usepackage{amssymb}

\usepackage{hyperref}
\hypersetup{
	colorlinks,
	linkcolor={blue!90!black},
	citecolor={blue!90!black},
	urlcolor={blue!90!black}
}

\usepackage{mathtools}
\DeclarePairedDelimiter\lr{\lparen}{\rparen}
\DeclarePairedDelimiter\Lr{\lbrack}{\rbrack}
\DeclarePairedDelimiter\LR{\lbrace}{\rbrace}
\DeclarePairedDelimiter\avg{\langle}{\rangle}

\DeclarePairedDelimiter\ket{\lvert}{\rangle}

\DeclarePairedDelimiterX{\matrixel}[3]{\langle}{\rangle}{#1 \delimsize\rvert #2 \delimsize\lvert #3}
\newcommand{\DxDy}{\iiint\limits_{-\infty}^{\infty}\!\!\mathrm{D}x_P\mathrm{D}x_Q\mathrm{D}x_B}

\newcommand{\Trpq}{\mathrm{Tr}_{pq}}
\newcommand{\Rkk}{\mathcal{R}_{kk'}}
\newcommand{\Ukk}{\mathcal{U}_{kk'}}
\newcommand{\JOBM}{\frac{J_0\beta}{M}}
\newcommand{\JBM}{\frac{J\beta}{M}}
\newcommand{\pkia}{p_{i\alpha}^{(k)}}
\newcommand{\pkja}{p_{j\alpha}^{(k)}}
\newcommand{\qkia}{q_{i\alpha}^{(k)}}
\newcommand{\qkja}{q_{j\alpha}^{(k)}}
\newcommand{\Ska}{\sum_{k\alpha}}
\newcommand{\Si}{\sum_i}
\usepackage{xcolor}

\newcommand{\e}{\mathrm{e}}

\journal{Physica A}

\begin{document}

\begin{frontmatter}

\title{Reentrant phase transitions involving glassy and superfluid orders in the random hopping Bose-Hubbard model}

\author[1]{Anna M. Piekarska}
\ead{a.piekarska@intibs.pl}
\author[1]{Tadeusz K. Kope\'{c}}
\affiliation[1]{organization={Institute of Low Temperature and Structure Research, Polish Academy of Sciences},
            addressline={Okólna~2}, 
            city={Wroc\l{}aw},
            postcode={50-422}, 
            country={Poland}}
            
\begin{abstract}
We study a system of strongly correlated bosons with off-diagonal disorder, i.e., randomness in the kinetic energy,
and find a family of reentrant phase transitions that occur as a function of the on-site interaction.
We model the system using the paradigmatic Bose-Hubbard Hamiltonian with a random hopping term
and solve it employing the replica trick and Trotter-Suzuki expansion known from quantum spin-glasses.
From subsequent numerical calculations, we find three distinct phase boundaries at which the reentrant transitions occur:
between glass and disordered phase, between superglass and superfluid ones, and between superfluid and disordered phases.
All three happen at temperatures slightly above critical temperatures of corresponding non-interacting systems.
When the emerging and disappearing order is glassy, this corresponds to the interplay of the thermal energy and the spread of hoppings.
When superfluidity is involved, thermal fluctuations must slightly overcome the mean hopping in turn for the reentrance to occur.
\end{abstract}

\end{frontmatter}
%
\section{Introduction}
Reentrant phase transitions are indeclinably arousing interest
and are found across a wide range of fields of study not limited to physics~\cite{Cladis1988_MCaLCINO165,Narayanan1994_PR249,Portz2021_JMB433}.
They occur when a particular phase, characterized by a specific order found in the state of the system,
disappears and later reappears as a single system parameter is changed.
Such behavior is counter-intuitive, as temperature and other system parameters typically act one-sidedly, building up or destroying a given type of order.
Observations of reentrant transitions date back to the 19th century~\cite{Demus1998_LCT8},
but new instances keep being discovered~\cite{Lake2021_PRB104} and are no less intriguing.

In physics, notably rich phase diagrams are found in highly correlated spin~\cite{Takayama1992_JPSJ61,Gong2014_PRL113} and many-body~\cite{Kopec1990_JoPCM2,Gurarie2009_PRB80} systems,
in which the incorporated disorder adds another level of complexity to the observed phenomena~\cite{Miranda2021_PRB104}.
Among these, strongly correlated disordered bosonic systems are of particular interest~\cite{Zapf2014_RoMP86}.
Next to the widely studied diagonally-disordered ones~\cite{Fisher1989_PRB40,Scalettar1991_PRL66,Weichman2008_PRB77,Gurarie2009_PRB80},
the off-diagonal disorder class offers a connection to the spin-glass~\cite{Sherrington1975_PRL35,book_Mezard1986} realm
--- the paradigmatic foundation for understanding disordered systems.
While the diagonal disorder, i.e., the one present in the potential energy, leads to the emergence of the amorphic Bose glass (BG) phase~\cite{Weichman2008_PRB77},
the off-diagonal one, meaning randomness of the kinetic energy,
has been recently shown~\cite{Piekarska2018_PRL120,Piekarska2020_JoSMTaE2020,Piekarska2022_PRB105} to result in a different kind of glassiness
involving complex phases of bosons in analogy to spin orientation in spin-glass systems.

The topic of reentrant phase transitions has been explored up to some degree in the context of both strongly correlated bosons and disordered systems.
The most basic example is the reentrance of the superfluid (SF) phase from the Mott insulator (MI) one
when changing the chemical potential (affecting the mean number of particles) in a system of lattice bosons.
However, it is easily understandable due to the specific character of the chemical potential
and the resulting quasi-periodicity of system properties as its function.
Other examples include the reentrance of MI from SF as the hopping integral~\cite{Elstner1999_PRB59},
or the on-site interaction~\cite{Kleinert2004_PRL93,CapogrossoSansone2007_PRB75,CapogrossoSansone2008_PRA77} is varied.
In the latter case, it has been shown that the intermediate values of the on-site interaction lead to the increase of the critical temperature.

In the diagonally-disordered interacting boson systems,
a reentrant behavior of the Bose glass was found with the inclusion of the superfluid phase~\cite{Hatano1995_JPSJ64,Gurarie2009_PRB80}.
There, changing the value of the on-site interaction moved the system from BG to SF and then to BG again.
In these systems, a notion of a reentrant superfluid is also in use~\cite{Soeyler2011_PRL107,Abreu2018_PRA98}, but with a different meaning --
there, it denotes a superfluid phase that emerges due to the disorder when the clean system is insulating.
The same feature of the phase diagram contains also a reentrance (of BG as a function of disorder) in the meaning used here, but it was not analyzed by the authors.
Note that these phenomena are qualitatively different from those studied here, as different types of glassines are involved.

In the random spin systems, a spin-glass phase occurring at temperatures lower than the ferromagnetic one is usually called reentrant~\cite{Binder1986_RoMP58}.
It is, however, not a typical reentrant phase, as it does not exist at higher temperatures.
The reason for using the term is that the high-temperature paramagnetic phase also lacks the ferromagnetic order.
Nevertheless, a genuine reentrant transition from the ferromagnetic to the paramagnetic phase was eventually found in random spin systems~\cite{Gingras1998_PRB57}.

Here, we report a family of reentrant phase transitions
as a function of the on-site interaction in an off-diagonally-disordered system of interacting bosons.
We find that the reentrant transitions between various phases
occur at temperatures slightly above the critical temperatures of corresponding non-interacting systems.
For the glassy ordering, this happens when the thermal energy is comparable to the disorder,
while for the superfluid one --- when it is comparable to the mean hopping.
The reentrant behavior can be found on three different phase boundaries:
between the glass (GL) and disordered (DI) phases, between the superglass (SG) and SF phases, and on the SF-DI boundary.
The last one is analogous to the one known from the non-disordered system of interacting bosons~\cite{Kleinert2004_PRL93},
yet we show its existence also in the presence of the disorder.
The other two have no analogues in the literature.
We present the phase diagrams showing the reentrant behavior
and characterize the phases using values of the order parameters along adequate cross-sections.

\section{Model}
We describe the system using the Bose-Hubbard Hamiltonian~\cite{Fisher1989_PRB40}
\begin{equation}
	H = -\sum_{i<j}J_{ij}\lr*{a_i^\dagger a_j+a_j^\dagger a_i}
	    + \frac{U}{2}\Si \hat{n}_i\lr*{\hat{n}_i-1} - \mu \Si \hat{n}_i,
\end{equation}
where $a_i$ ($a_i^\dagger$) annihilates (creates) a boson at site $i$,
and $\hat{n}_i=a_i^\dagger a_i$ are the corresponding particle number operators.
In this Hamiltonian, $\mu$ denotes the chemical potential, $U$ is the on-site interaction,
while $J_{ij}$ are hopping integrals defined as independent Gaussian-distributed random variables
with the mean $J_{0}/N$ and variance $J^{2}/N$, as has been used in the field of spin glasses~\cite{Sherrington1975_PRL35}.
Also in agreement with spin-glass models, the hopping is active between each pair of sites, which makes it effectively long-range.
The experimental realization of such a setting has been suggested in Ref.~\cite{Piekarska2018_PRL120},
and we believe it can be feasible with currently available optical-lattice quantum simulators~\cite{Cataliotti2001_S293}.

In the following, before discussing the results, we briefly outline the derivation,
which follows the scheme of the Sherrington-Kirkpatrick model of spin glasses~\cite{Sherrington1975_PRL35}
while the full derivation and description of the current case can be found in Ref.~\cite{Piekarska2022_PRB105}.

\section{Methods}
We start from the free energy, $F = -(1/\beta) \ln Z$, where $Z=\mathrm{Tr} \exp(-\beta H)$ is the partition function, $\beta=1/(k_{\mathrm{B}}T)$, and we work with $k_{\mathrm{B}}=1$.
We average it over the disorder analytically and apply the replica~\cite{Sherrington1975_PRL35} and Trotter-Suzuki methods~\cite{Suzuki1985_PLA113}.
In the replica trick, one replaces the logarithm according to
\begin{equation}
	\ln Z = \lim_{n\to 0} \frac{1}{n}\lr*{Z^n-1},
\end{equation}
which makes the averaging over the disorder possible, but at the cost of introducing the replica space.
Next, the Trotter-Suzuki method allows one to deal with an expression of the form $\e^{H_A+H_B}$, where $H_A$ and $H_B$ are operators that do not commute.
One does so by introducing a new time-like dimension $M$,
\begin{equation}
	\exp \lr*{-\beta (H_{A}+H_{B})} = \lim_{M\to\infty} \Lr*{\exp \lr*{-\beta H_A}\exp \lr*{-\beta H_B}}^M.
\end{equation}
This is followed by inserting a summation over projectors onto a complete set of eigenstates that allow replacing the operators with corresponding eigenvalues.
Such a procedure maps a quantum model onto a corresponding classical one with an additional dimension.
At this point, we integrate over the distributions of the variables $J_{ij}$ and arrive at the effective replicated partition function
\begin{equation}\label{eq:Zn}\begin{split}
	\Lr*{Z^n}_{J} = \Trpq \mathcal{M}_{pq}
	\prod_{i < j} \exp \LR[\Bigg]{&
		\frac{J^2\beta^2}{2M^2N}\Lr*{
			\Ska\lr*{\pkia\pkja + \qkia\qkja}
		}^2\\
		&+\frac{\beta J_0}{MN} \Ska\lr*{
			\pkia\pkja + \qkia\qkja
		}
	},
\end{split}\end{equation}
where $\Lr*{\cdot}_{J}$ denotes the average over the distribution describing the disorder,
$\alpha = 1, \ldots, n$ are the replica indices,
and $k = 1, \ldots, M$ are the Trotter indices of time-like character.
Moreover, variables $\pkia$ and $\qkia$ are the eigenvalues of the $\hat{P}=i\lr{a^\dagger-a}/\sqrt{2}$
and $\hat{Q}= \lr{a^\dagger+a}/\sqrt{2}$ operators, corresponding to eigenvectors $\ket{\pkia}$ and $\ket{\qkia}$, accordingly.
The trace runs over all possible values of $\pkia$ and $\qkia$,
and $\mathcal{M}_{pq}$ is a product of matrix elements given by
\begin{equation}
	\mathcal{M}_{pq} = \prod_i \mathcal{M}_{pq}^{(i)}
	=\prod_{ik\alpha}
		\matrixel*{\pkia}{e^{-\frac{\beta H_n}{2M}}}{\qkia}
		\matrixel*{\qkia}{e^{-\frac{\beta H_n}{2M}}}{p_{i\alpha}^{(k+1)}}.
\end{equation}
Here,
\begin{equation}
	H_n = \frac{U}{2}\sum_{i\alpha}\hat{n}_{i\alpha}^2-\lr*{\mu+\frac{U}{2}}\sum_{i\alpha}\hat{n}_{i\alpha}
\end{equation}
is the last indication of the quantum nature of the system in an otherwise classical effective problem.

We use the Hubbard-Stratonovich transformation to decouple site-mixing terms in Eq.~\ref{eq:Zn}.
This yields a new set of fields, $\Rkk$, $\Ukk$, $\varDelta$, $q$, and $u$, which couple to the decoupled terms.
Next, in the thermodynamic limit we use the saddle point method.
In order to take the limit $n\to 0$, we apply the Hubbard-Stratonovich transformation again,
this time decoupling the replica-mixing terms and introducing auxillary integration variables $x_{P}$, $x_{Q}$ and $x_{B}$.
After additionally assuming symmetry in the replica space, we finally take the $n\to 0$ limit
to arrive at the effective free energy and a system of self-consistent equations.
The free energy reads
\begin{multline}
	\mathcal{F} =
		2\lr[\Big]{\JBM}^{2}\sum_{kk'}\lr*{\Rkk^2+\Ukk^2} + J_{0}\beta\varDelta^2 - 2\lr[\Big]{\frac{J\beta}{M}}^{2}\lr*{q^2+u^2}\\
		- \DxDy \ln \Trpq \exp \lr*{-\beta \mathcal{H}},
\end{multline}
where we have defined $\mathrm{D}x \equiv \mathrm{d}x\exp(-x^2)$
and the corresponding effective Hamiltonian is
\begin{multline}\label{eq:effham}
	-\beta\mathcal{H} =
		2\lr[\Big]{\JBM}^{2}\sum_{kk'}\Lr[\Big]{
			(\Rkk-q)\lr*{p_{k}p_{k'} + q_{k}q_{k'}}
			+2(\Ukk-u)p_{k}q_{k'}
		} + 
		\varDelta\JOBM \sum_{k}\lr*{p_{k}+q_{k}} +\\ 
		+\frac{2J\beta}{M}\sum_{k}\lr*{ x_{B}\sqrt{\frac{u}{2}}(p_{k}+q_{k}) + \sqrt{\frac{q-u}{2}}(x_{P}p_{k}+x_{Q}q_{k}) }
		+\ln \mathcal{M}_{pq}.
\end{multline}
The self-consistent equations are
\begin{subequations}
\begin{eqnarray}
	\label{eq:selfcon_q} q &=& \DxDy \avg{p_k}^2,\\
	\label{eq:selfcon_D} \varDelta &=& \DxDy \avg{p_k},\\
	\label{eq:selfcon_u} u &=& \DxDy \avg{p_k}\avg{q_k},\\
	\label{eq:selfcon_R} \Rkk &=& \DxDy \avg{p_kp_{k'}},\\
	\Ukk &=& \DxDy \avg{p_kq_{k'}},
\end{eqnarray}
\end{subequations}
where
\begin{equation}
	\avg{A} = \frac{\Trpq A \exp(-\beta\mathcal{H})}{\Trpq \exp(-\beta\mathcal{H})},
\end{equation}
are thermal averages taken with the effective Hamiltonian \eqref{eq:effham}.
The first two quantities, $q$ and $\varDelta$, coincide with the order parameters
for the glassy (Edwards-Anderson order parameter, see Ref.~\cite{Edwards1975_JoPFMP5}) and superfluid~\cite{Fisher1989_PRB40} orders, respectively.
The third value, $u$, is secondary to $\varDelta$ and, as has been shown in Ref.~\cite{Piekarska2022_PRB105},
is equal to $q$ in the superfluid phase, while its value is $0 < u < q$ in the superglass phase, which combines both kinds of order.
Finally, $\Rkk$ and $\Ukk$ are dynamic self-correlations discussed in Ref.~\cite{Piekarska2020_JoSMTaE2020}.

\section{Recognizing the phases}
In the system, we expect two kinds of ordering --- glassy and superfluid,
characterized by the order parameters $q$ and $\varDelta$, respectively.
Their presence and interplay result in the emergence of four phases: disordered, containing neither of these orderings,%
\footnote{
At the lowest temperatures this phase is the MI phase, at higher ones it may be a normal liquid.
However, we do not aim to distinguish the phases within this class here.}
glass and superfluid phases containing glass and superfluid orders, respectively,
as well as the superglass phase, containing both orders simultaneously.
To distinguish the phases, we derive and then numerically evaluate Landau-theory conditions,
which allows us to differentiate between DI, GL, and SF+SG phases.
In order to detect the glassy order on top of the superfluid one, and thus identify the SG phase from the SF one,
we use the instability of the replica-symmetric solution as a signature of glassiness.
The appropriate condition can be found in Ref.~\cite{Piekarska2022_JSM2022}.
These conditions give us a clear distinction between the phases,
nevertheless, we can also characterize the phases by directly checking the values of order parameters.

\section{Results and discussion}
\begin{figure}[tbp]
	\centering
	\includegraphics[width=0.64\columnwidth]{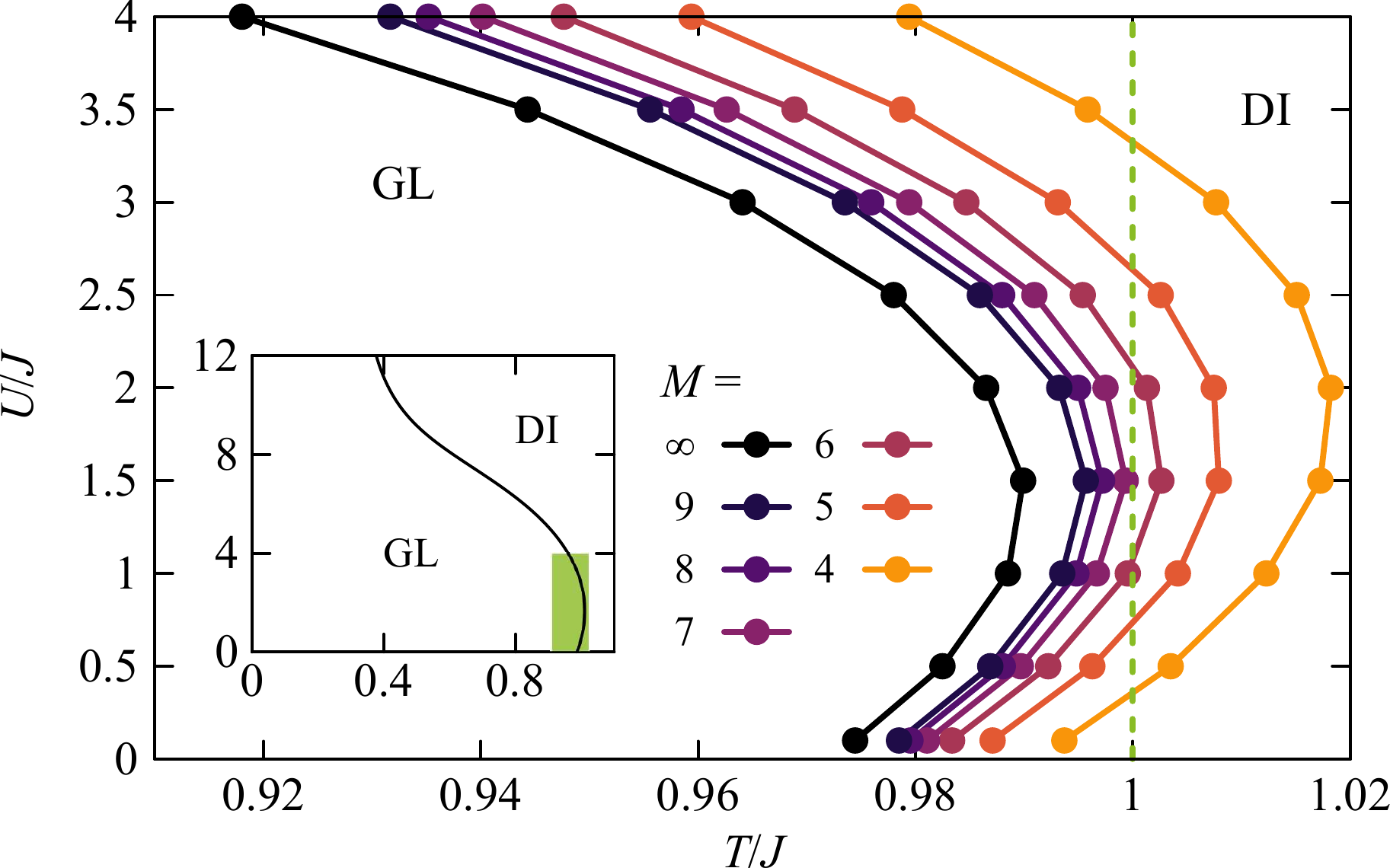}
	\caption{Phase diagram in variables $T/J$--$U/J$ at constant $\mu/U = 0.4$ and $J_0/J = 0.8$.
		The DI-GL critical lines obtained with finite values of $M=4$--$9$ (yellow-purple points)
		as well as their $M\to\infty$ extrapolation (black points) are shown.
		The vertical line marks the location of the cross-section used in the subsequent figure.
		The inset contains the zoomed out phase diagram at $M=5$ with the area of the main plot marked with a green rectangle.
		On both plots, the phases are marked.}
	\label{fig:J08}
\end{figure}
\begin{figure}[tbp]
	\centering
	\includegraphics[width=0.82\columnwidth]{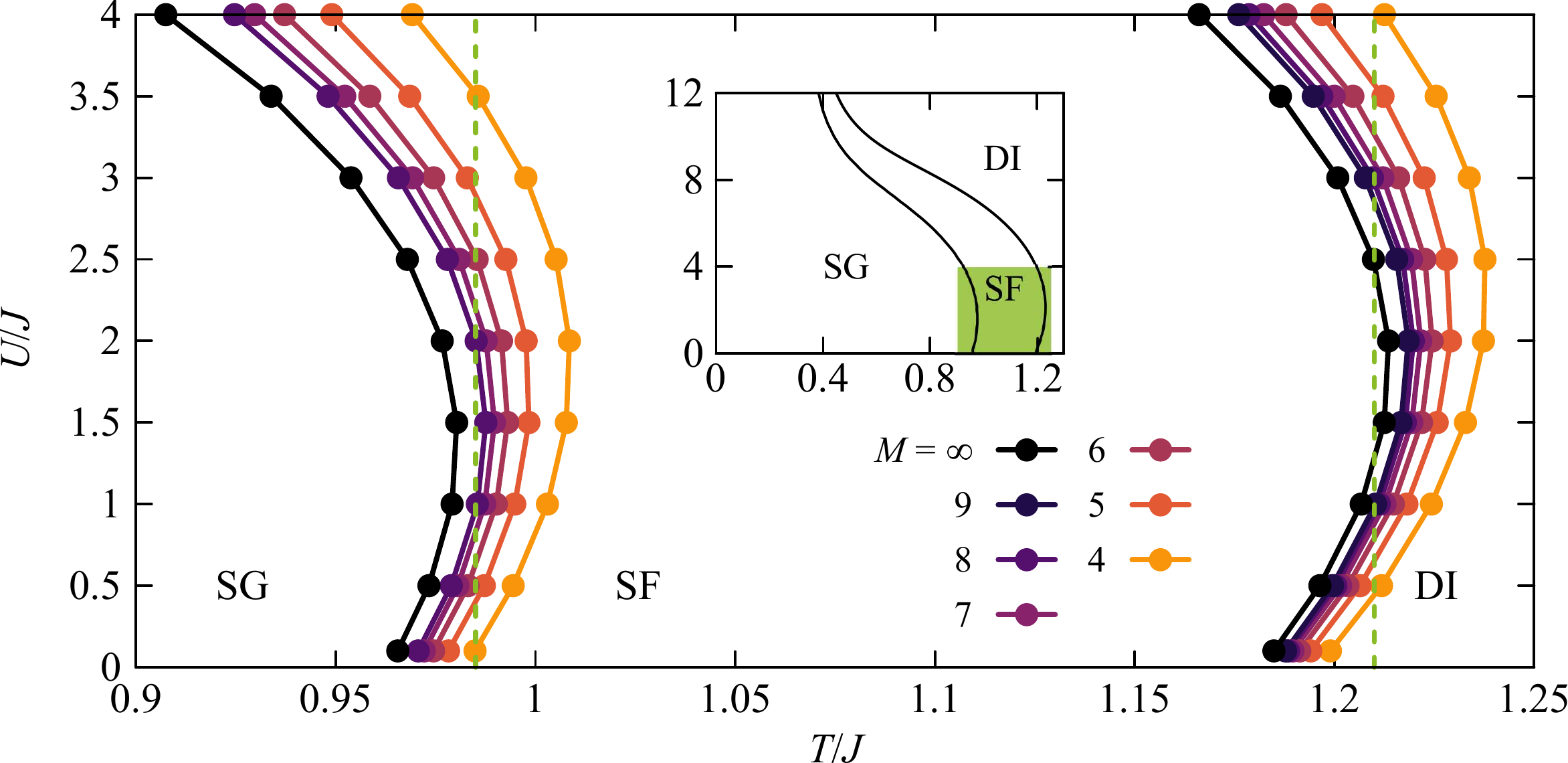}
	\caption{Phase diagram in variables $T/J$--$U/J$ at constant $\mu/U = 0.4$ and $J_0/J = 1.2$.
		The DI-SF and SF-SG critical lines obtained with finite values of $M=4$--$9$ (yellow-purple points)
		as well as their $M\to\infty$ extrapolation (black points) are shown.
		The vertical lines mark the locations of the cross-sections used in a subsequent figure.
		The inset contains the zoomed out phase diagram at $M=5$ with the area of the main plot marked with a green rectangle.
		On both plots, the phases are marked.}
	\label{fig:J12}
\end{figure}

We find that there are multiple kinds of reentrant phase transitions along the $U/J$ axis,
meaning that there is a range of values of $U/J$, i.e., interaction strengths, for which the critical temperature of various transitions is higher than the one from the noninteracting, $U=0$, system.
The first one, shown in Fig.~\ref{fig:J08}, occurs at $J_{0} < J$, where the glass phase intervenes between two areas of the disordered one.
In the figure, we plot the finite-$M$ DI-GL critical lines as well as their $M\to\infty$ extrapolation~\cite{Suzuki1985_PLA113}.
As the main plot covers narrow ranges of the parameters,
in the inset, we present the phase diagram in their broader range (at $M=5$) to place the reentrant part in the bigger picture.
One may compare this result to quantum spin glass systems~\cite{Usadel1986_SSC58,Buettner1991_ZfPBCM83}, where a very similar phase boundary between the paramagnetic and glassy regions was found, but this kind of reentrant behavior was not seen.
As the used model is equivalent to the one used for spin glass systems,
we establish that the on-site repulsion $U$ acts in a qualitatively distinct way from the transverse field in spin glasses.
While the field works one-sidedly and only destroys the order,
there is a specific range of temperatures in which the interactions turn out to be necessary to stabilize the ordered phase.
Only after reaching some higher critical value do the interactions destroy the order.
One might notice that the transition occurs around the value of $T/J\approx 1$.
This is not a coincidence and we have confirmed that the $U=0$ DI-GL transition
happens at $T \approx J$ for other valid values of $J_{0}/J$ as well (the transition exists only at $J_{0}<J$).

The other two kinds of reentrant phase transitions occur at $J_{0} > J$ and are presented using the value of $J_{0}/J = 1.2$ in Fig.~\ref{fig:J12}.
The first one takes place at $T/J\approx 1.2$, where the SF phase appears between two areas of the disordered one.
This kind of reentrant behavior has been seen in a non-disordered bosonic model in Ref.~\cite{Kleinert2004_PRL93}.
Here we confirm its occurrence also in a system that is additionally disturbed by the disorder.
The other reentrant behavior takes place at $T/J\approx 1.0$, where we observe a SF phase both above and below the SG one.
As in the case of the previous figure, we plot the finite-$M$ critical lines and their $M\to\infty$ extrapolation.
Again, we accompany the plot with one with a broader range of parameters.
Similarly as in the previous case, the DI-SF transition takes place at thermal energies comparable to $J$.
The critical value of $T/J\approx 1.2$ for the SF-SG line is not a coincidence either
--- from the analysis of the phase transition at other values of $J_{0}/J$ (the transitions exist at $J_{0}>J$),
we have established that it always happens at $T \approx J_{0}$, thus when thermal fluctuations are comparable to the mean hopping.

\begin{figure}[tbp]
	\centering
	\includegraphics[width=0.90\columnwidth]{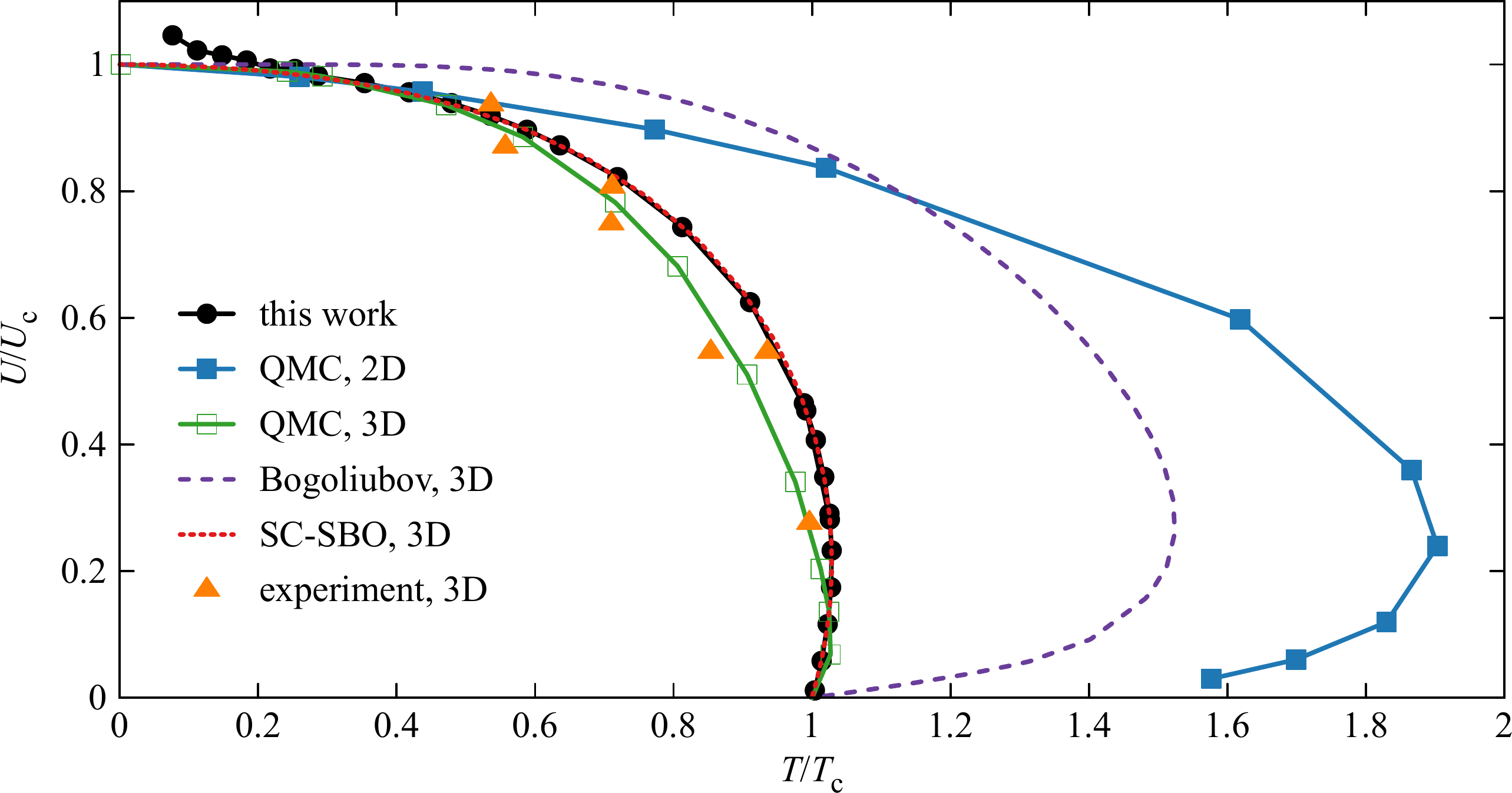}
	\caption{Phase diagram in variables $T/T_{\mathrm{c}}$--$U/U_{\mathrm{c}}$:
		a comparison between the current formalism (with disorder) and analogous results for the clean system.
		The value $T_{\mathrm{c}}$ denotes the critical temperature of the non-interacting system,
		while the value $U_{\mathrm{c}}$ corresponds to the critical value of the on-site interaction at $T=0$.
		The line from the current formalism (at $M\to\infty$) is drawn with filled black circles.
		The currently calculated points are accompanied by our previous results~\cite{Piekarska2022_PRB105} in the low-temperature regime.
		The critical lines drawn with squares were obtained from quantum Monte Carlo (QMC) simulations
		in 2D (full blue squares, Ref~\cite{CapogrossoSansone2008_PRA77}) and 3D (empty green squares, Ref~\cite{CapogrossoSansone2007_PRB75}).
		The purple dashed line is an analytical solution obtained through the Bogoliubov approximation
		and evaluated using the hopping expansion~\cite{Kleinert2004_PRL93}.
		The red dotted line was obtained through self-consistent standard basis operator (SC-SBO) approach~\cite{Sajna2015_PRA92}.
		The orange triangles correspond to experimentally obtained phase boundaries~\cite{Trotzky2010_NP6}.
		Note that since we do not know the exact value of $U_{\mathrm{c}}$ or $T_{\mathrm{c}}$ in some cases,
		the quantitative relative scaling between the curves may not be accurate.}
	\label{fig:lit}
\end{figure}

Reentrant behavior on the DI-SF critical line has been seen in non-disordered systems.
Since this field is more explored, the transition has been found using various methods and with very high precision.
Most of the analyses have been performed on finite-dimensional systems,
which can not be directly compared to our results quantitatively,
but nevertheless, a qualitative comparison is possible.
In Fig.~\ref{fig:lit}, we present the temperature-dependent critical lines
obtained from the current formalism (black circles) along with those obtained through other methods for the clean system:
quantum Monte Carlo simulations in 2D~\cite{CapogrossoSansone2008_PRA77} and 3D~\cite{CapogrossoSansone2007_PRB75};
analytical solution obtained through the Bogoliubov approximation applied to a 3D system and evaluated using the hopping expansion~\cite{Kleinert2004_PRL93};
self-consistent standard basis operator approach in which the random phase approximation for the 3D Bose-Hubbard Hamiltonian is used~\cite{Sajna2015_PRA92};
as well as the experimental data from a 3D system of ultracold bosons in an optical lattice~\cite{Trotzky2010_NP6}.
All these lines are qualitatively similar -- there is a reentrant phase transition
at temperatures slightly above the critical temperature of the non-interacting system.

\begin{figure}[tbp]
	\centering
	\includegraphics[width=0.53\columnwidth]{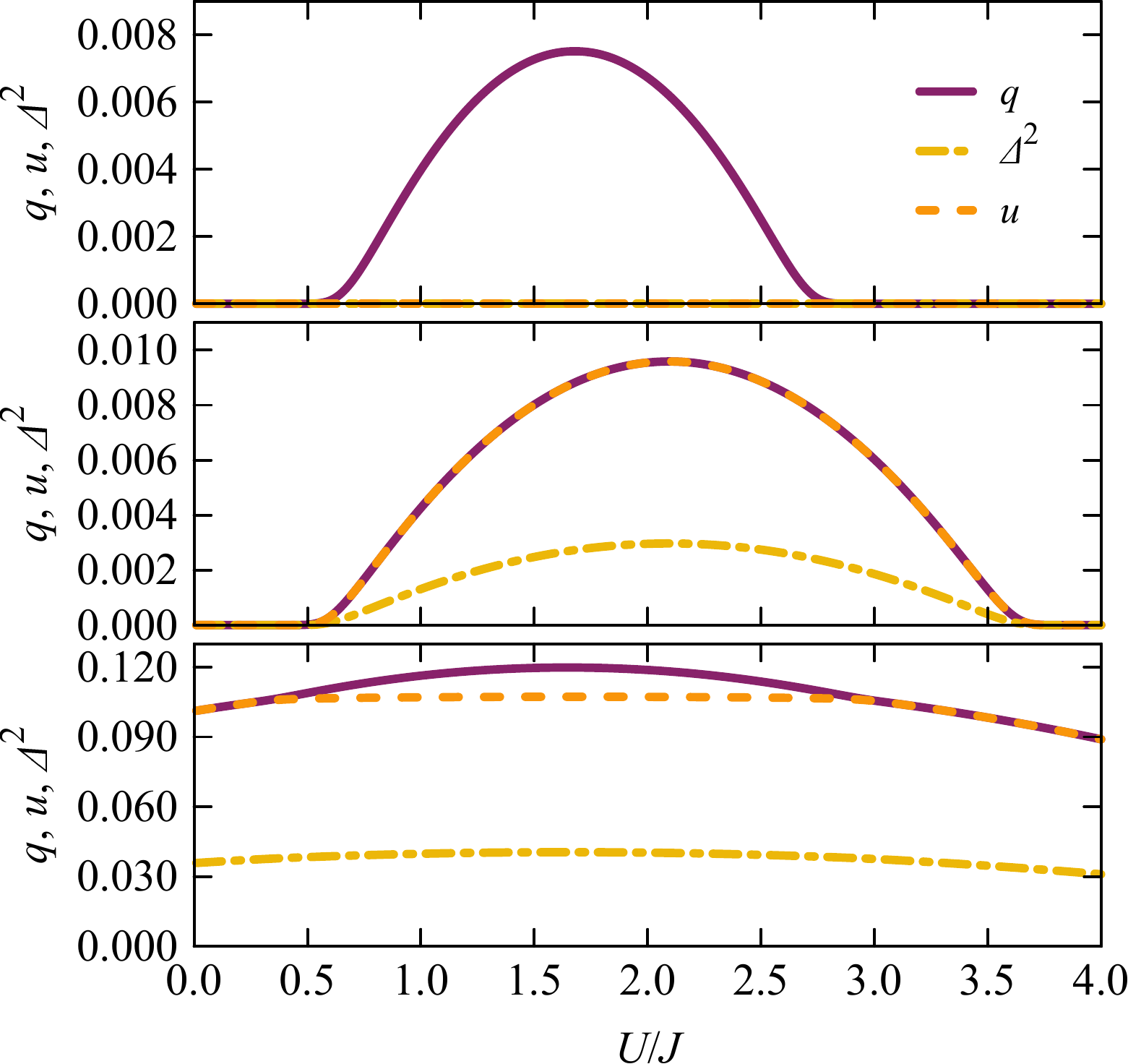}
	\caption{Values of the order parameters obtained with $M=5$ along the vertical cross-sections marked on (a) Fig.~\ref{fig:J08} and (b,c) Fig.~\ref{fig:J12},
	i.e., obtained at constant $\mu/U = 0.4$ and (a) $T/J = 1.0$, $J_0/J = 0.8$,
	(b) $T/J = 0.985$, $J_0/J = 1.2$ and (c) $T/J = 1.21$, $J_0/J = 1.2$.
	$q$ and $\varDelta$ are glass and superfluid order parameters, respectively,
	while $u$ is a value secondary to $\varDelta$.}
	\label{fig:cuts}
\end{figure}

We now turn our attention to the cross-section of the phase diagram across which the reentrant behavior happens.
In Fig.~\ref{fig:cuts}, we present three such cross-sections corresponding to three distinct critical lines presented above.
The exact positions of the cross-sections are marked in Fig.~\ref{fig:J08} and Fig.~\ref{fig:J12} by dashed vertical lines.
Along these cross-sections, we present the values of the order parameters $q$ and $\varDelta$, as well as the auxiliary variable $u$.
The curves are obtained with a finite value of $M=5$, but their qualitative behavior should stay the same with $M\to\infty$,
what is, however, not feasible to obtain numerically.
On all three cross-sections, we observe a qualitative change of behavior twice ---
first, from a less ordered phase to a more ordered one, and the opposite transition later.
In the panel (a), we deal with the appearance of a non-zero value of the glass order parameter $q$
at around the value $U/J\approx 2.7$ and then its disappearance at $U/J\approx 0.7$.
In the panel (b), at $U/J\approx 3.4$, both $q$ and $\varDelta$ become non-zero,
signaling the presence of the superfluid phase to later vanish around $U/J\approx 0.8$.
Finally, in panel (c), the values of $q$ and $u$ become distinct at $U/J\approx 2.9$,
which we mentioned earlier as the signature of the superglass phase, and become equal again at $U/J\approx 0.4$.

\FloatBarrier
\section{Conclusions and outlook}
We have presented a family of reentrant phase transitions occurring along the $U/J$ axis in a Bose-Hubbard model with random hopping integrals.
These transitions take place on three distinct phase boundaries: glassy to disordered, superfluid to disordered, and superglass to superfluid.
We have found that at intermediate values of the on-site repulsion,
the critical temperatures are slightly higher than those of either their non-interacting or strongly interacting counterparts.
This suggests that the on-site repulsion can stabilize the ordered phase above the non-interacting temperature threshold.
However, the question of \emph{why} it happens remains open.
Providing an unambiguous answer can be difficult.
In many cases, one usually limits the study to establishing the presence of reentrant behavior.
It is possible that numerical studies in finite dimensional systems, providing insight into the microscopic properties of the phases,
could bring us closer to understanding the observed behavior.

\section*{Acknowledgments}
This work was supported by the Polish National Science Centre under Grant No\linebreak2018/31/N/ST3/03600.
Calculations have been partially carried out using resources provided by Wroclaw Centre for Networking and Supercomputing (http://wcss.pl), grant No. 449.


\end{document}